\documentclass[11pt]{article}
\usepackage{times}
\usepackage{geometry}
\usepackage[dvipsnames]{xcolor}
\usepackage[utf8]{inputenc}
\usepackage{enumitem,amssymb}
\usepackage{graphicx}
\usepackage{fancyhdr}
\usepackage{aas_macros}
\usepackage{mdframed} 
\usepackage{dblfnote}
\usepackage{multicol}

\usepackage{hyperref}
\usepackage{scrextend}

\hypersetup{
    bookmarks=true,         
    unicode=false,          
    pdftoolbar=true,        
    pdfmenubar=true,        
    pdffitwindow=false,     
    pdftitle={Certificate},    
    pdfauthor={Christopher D. Matzner et al.},     
    pdfsubject={Low-Carbon Astronomy},   
    pdfcreator={Chrsistopher D. Matzner},   
    pdfproducer={},  
    pdfkeywords={Climate Change}{Professional Practices}{Astronomy}{Sustainability}, 
    pdfnewwindow=true,      
    colorlinks=false,       
    linkcolor=red,          
    citecolor=green,        
    filecolor=magenta,      
    urlcolor=cyan           
}

\mdfdefinestyle{theoremstyle}{
innertopmargin=\topskip,}
\mdtheorem[style=theoremstyle]{lrptextbox}{}

\newcommand{\cotwo}{CO$_2$}
\newcommand{\cdm}{}

\newcommand{\symfootnote}[1]{%
\let\oldthefootnote=\thefootnote%
\stepcounter{mpfootnote}%
\addtocounter{footnote}{-1}%
\renewcommand{\thefootnote}{\alph{mpfootnote}}%
\footnote{#1}%
\let\thefootnote=\oldthefootnote%
}

\title{Astronomy in a Low-Carbon Future \\ (A White Paper prepared for the Canadian {\em Long Range Plan} 2020)} 

\begin{document}

\maketitle













\centerline{\bf\underline{\large{ Executive summary}}} 
\vspace{0.1in}

The consequences of climate change are {developing} into a vast and unparalleled crisis, which is planetary in scale and yet very human in its causes and impacts.  Faced with this truth, inaction and delay are neither ethical nor wise: every field of human activity, astronomy included, must take urgent steps to mitigate the crisis and avoid the worst potential outcomes, while adapting to those that are inevitable.  We outline a strategy for climate-responsible scholarship and communication that we hope will resonate with the astronomical community's aspirations for a stable future in which there is room for curiosity and discovery.

In light of the crisis, greenhouse gas emissions must be understood as significant research costs to be justified, budgeted, and rationed.  Air travel is a major contributor.  We must remove incentives in tenure and promotions for profligate travel, codify low-carbon practices, and reward efforts to minimize emissions while maximizing research.  Any essential travel emissions must be fully offset, with carbon offsets that are reimbursable if not covered by the institution.  

The frequency, timing, and locations of conferences and colloquia should be planned to minimize air travel, with remote participation encouraged using online platforms.   Conferences should favour lowest-impact food options by default, taking steps to minimize food waste (and other waste as well).

Infrastructure must be planned with climate impacts in mind for both construction and operation.   For buildings, this means pushing for designs and materials at the forefront of sustainable technology.  For observatories and computing facilities, it includes justifying and thoroughly vetting the emissions footprint at the proposal stage and fully considering low-impact options, such as cloud computing.

In education and outreach, we must strive to be reliable sources of scientific information on climate change, prepared to connect this to astrophysics and ready to direct curious people to authoritative sources.

We recommended that CASCA support this transformation by establishing a sustainability committee (or office) to be tasked with conducting a community-wide survey of our greenhouse gas emissions and developing a decarbonization road map for the discipline. This committee would host sustainability forums and provide resources to support each of our other recommendations, especially climate-related materials and references for education and outreach.

We also recommend that CASCA push for climate-responsible governance.  Governments, granting agencies, and universities must implement climate-responsible policies, like rationing the carbon budget of the entire discipline,  using climate impacts to evaluate proposals, permitting the use of funds for offsets, and allowing climate-friendly but more costly travel options.   Moreover, these entities must ultimately ration emissions to ensure that targets are met.

To have an impact beyond the bounds of professional astronomy, we encourage astronomers to press their institutions to divest from fossil fuel extraction and its financing.

Finally, we recommend that climate responsibility be an explicit priority in the 2020 {\em Long Range Plan}, and that it be used to guide all aspects of our profession, from our research and education practices to our infrastructure.

\eject

\centerline{\Large{Astronomy in a Low-Carbon Future}}
\vspace{0.2in}
\author{Christopher D. Matzner,\symfootnote{University of Toronto} 
Nicolas B. Cowan,\symfootnote{McGill University} 
Ren\'e Doyon,\symfootnote{Director, Inst.\ for Research on Exoplanets}$^,$\symfootnote{Universit\'e de Montreal} 
Vincent H\'enault-Brunet,\symfootnote{Saint Mary’s University} 
David Lafreni\`ere,$^{\rm d}$
Martine Lokken,$^{\rm a}$ 
Peter G.~Martin,$^{\rm a,}$\symfootnote{CITA} 
Sharon Morsink,\symfootnote{University of Alberta} 
Magdalen Nomandeau,\symfootnote{University of New Brunswick} 
Nathalie Ouellette,$^{\rm d}$
Mubdi Rahman,$^{\rm a,}$\symfootnote{Dunlap Institute} 
Joel Roediger,\symfootnote{NRC Herzberg} 
James Taylor,\symfootnote{University of Waterloo} 
Rob Thacker,$^{\rm e}$
and 
Marten van Kerkwijk$^{\rm a}$  
}

\begin{abstract}
    
The global climate crisis poses new risks to humanity, and with them, new challenges to the practices of professional astronomy.  Avoiding the more catastrophic consequences of global warming by more than 1.5 degrees requires an immediate reduction of greenhouse gas emissions. 
According to the 2018 United Nations Intergovernmental Panel report, this will necessitate a 45\% reduction of emissions by 2030 and net-zero emissions by 2050.  Efforts are required at all levels, from the individual to the governmental, and every discipline must find ways to achieve these goals.  This will be especially difficult for astronomy with its significant reliance on conference and research travel, among other impacts.  However, our long-range planning exercises provide the means to coordinate our response on a variety of levels. We have the opportunity to lead by example, rising to the challenge rather than reacting to external constraints. 

We explore how astronomy can meet the challenge of a changing climate in clear and responsible ways, such as how we set expectations (for ourselves, our institutions, and our granting agencies) around scientific travel, the organization of conferences, and the design of our infrastructure.    We also emphasize our role as reliable communicators of scientific information on a problem that is both human and planetary in scale.  

\end{abstract}

\section{Background: the global climate crisis 
}\label{S:Background} 


As of September 2019 humans have emitted vast quantities of greenhouse gases (GHGs), most notably about 1.7 teratons$^1$ of \cotwo\ -- enough to raise the atmospheric levels 46\% above pre-industrial levels (280 ppm to 410 ppm),$^{2,3}$ half of this since 1990. 
Canada's emissions have been especially intense: at 15.6 tons \cotwo/yr per capita as of 2017,$^1$ more than triple the global average, it is in the top five large nations by this measure.
With secondary GHGs like methane and fast climate feedbacks (e.g., higher temperature leading to increased water vapour), the net effect has been global warming by about 1$^\circ$C since 1850,$^4$ with almost an additional half-degree of warming locked in$^5$ (on a four-decade delay due to ocean heat capacity).  Future warming predictions range from 1.5$^\circ$\,C to over 3.5$^\circ$\,C above pre-industrial levels,$^{6,7}$ depending on what fraction of fossil reserves are consumed; however paleoclimate studies$^8$ and recent modeling$^9$ suggest greater warming due to slow positive feedbacks (e.g., thawing permafrost releasing GHGs).  
As warming is enhanced near the poles,  Canada warms at twice the global average.$^{10}$ 

Warming induces a cascade of effects for humans and the natural world that have led to widespread calls for action, including increasingly urgent reports and warnings from the United Nations$^{6,7,11,12}$ and from governmental bodies (e.g., of Canada,$^{10}$  the US,$^{13,14}$ and Australia.$^{15}$ As of this writing, 1081 jurisdictions representing 266 million people in 20 countries have declared a
climate emergency.$^{16}$ 
We briefly review the reasons for this urgency. 

\noindent{\bf Melting ice, sea level rise:} A direct consequence of warming is the loss of land ice, and the associated rise in ocean levels (enhanced by thermal expansion).  
Arctic land ice melt has tripled since 1986, to a rate comparable to the flow of the Mississippi.$^{17}$ In Antarctica, recent rapid melting significantly exceeded the Arctic's and there is a likelihood of ice sheet destabilization.
Sea ice,$^{18}$ permafrost,$^{19}$ and glaciers$^{20}$ have all receded drastically, with positive feedbacks arising from decreased albedo and stored carbon.$^{21}$ 
Sea levels are rising 3.3 cm/decade now. This could result is a rise of anywhere between 0.2\, m and 2\,m by 2100,$^{22}$ and eventually $\sim$16\,m if we fail in our 1.5$^\circ$ goal, and see 2-3\,$^\circ$C warming instead.$^{23}$
As a result, low-lying island nations face imminent submergence;$^{24}$ glacier-fed watersheds and farmlands for several billion people are threatened;$^{25}$ and permafrost melting$^{26}$  undermines access and infrastructure$^{27}$ in Arctic regions of Canada, Scandinavia, Siberia, and Alaska. 

\noindent{\bf Natural disasters:}  
{In a warming climate, severe heat waves} become more frequent.$^{28}$ While the global mean atmospheric humidity rises in accordance with the  Clausius-Clapeyron relation, regional climatology becomes more ``summer-like'', so {droughts} are compounded in many regions,$^{29}$ with a quarter of the human population now at risk of drought.$^{30}$  { Wildfires} increase correspondingly,$^{31}$ now extending into the Arctic and throughout the year.  At the same time, { major storms} intensify due to increased evaporation,$^{32}$ leading (along with sea level rise) to dramatically more frequent and less predictable floods. Such trends have inspired warnings that multiple major disasters will coincide$^{33}$ and become uninsurable.$^{34}$
Costs accelerate rapidly above the 1.5$^\circ$\,C goal, approaching a quadrillion dollars (in 2019 CAD) over the next century if warming reaches 3$^\circ$.$^{35}$

\noindent{\bf Extinctions and threats to the biosphere:}
Climate change is one of many threats to the natural environment,$^{36}$  along with factors like rapid overdevelopment, deforestation, habitat restriction, overfishing, and pollution.  The result is a biodiversity crisis:
with 1.5-2 $^{\circ}$C of warming, the territorial ranges for a majority of terrestrial species will profoundly shrink, putting 5\% (10$^6$) at risk of extinction.  Another $\sim$10$^6$ species in marine environments are at risk from effects like the loss of coral reef habitats,$^{37}$ under threat from acidification$^{38}$ (a direct consequence of dissolved \cotwo), increased water temperatures, and the associated drop in dissolved O$_2$.  There has already been a sustained population decline across species;$^{39}$ the 29\% decline in North American bird populations since 1970 is a case in point.$^{40}$  All told, Earth's sixth mass extinction appears to be well underway.$^{41}$


\noindent{\bf Public health, food security:}
A changing climate threatens health, safety, and food security for large fractions of Earth's population, as outlined in  reports and declarations by major public health organizations.$^{42-47}$ Climate-induced spreading of diseases (Malaria, Dengue, Zika, Lyme, etc., due to expanding vector ranges), toxic algal blooms, air pollution (ozone, pollen, particulates from fossil fuel use and wildfires), direct heat stress, food and water insecurity,  extreme weather events, and psychological strain all contribute to the health crisis.  At the same time, global changes in temperatures, weather, hydroclimate, and insects are destabilizing food production,$^{48}$  while climate-related threats to fisheries create food insecurity for the billion people who rely on them. (In an unfortunate feedback, land clearing and land use for agriculture, especially cattle, cause major greenhouse gas emissions.) 

\noindent{\bf Global inequality, migration, and conflict:} Climate change is an important driver of global inequalities. Many of the poorest populations are located in (semi-)tropical regions that are much more vulnerable to the impacts of climate change,$^{49}$ 
a problem  compounded by fewer resources for responding to climate-related disasters. This inequality is already visible: in 2000-04, the fraction of the population affected by climate-related disasters was 80 times higher in `developing' countries than in the OECD.$^{50}$
As stated in one report of the United Nations Development Programme,$^{51}$ ``the rich contribute more [to environmental problems] but the poor bear the brunt in loss of lives and livelihoods.''
Developed countries have emitted the overwhelming majority of \cotwo; even now, North American per capita emissions  
are 50 times higher than those of the least developed countries and 20 times higher than those of Sub-Saharan Africa.$^{52}$  Populations severely exposed to climate risks would double if the 1.5$^\circ$ target were missed by 0.5$^\circ$, and would quadruple at  3$^\circ$.$^{53}$
This state of affairs is expected to create as many as 140 million climate refugees worldwide by mid-century.$^{54}$  Already, climate change is a primary driver of mass migrations in Central America, Syria, and elsewhere.  Unsurprisingly, therefore, it has been identified as a major and intensifying cause of conflict worldwide.$^{14,55}$ In Canada as elsewhere, climate change will hit Indigenous populations especially hard due to loss of opportunities to practise culturally and economically important activities, as well as loss of culturally significant sites, along with other stresses arising in historical and ongoing injustices -- with the subsequent loss of inter-generational cohesion and cultural integrity.$^{56-58}$

{\noindent \bf International commitments:}
In 2015, within the United Nations Framework Convention on Climate Change, 195 countries negotiated the `Paris Agreement' (ratified by Canada in 2016) pledging emissions reductions enough to hold the rise of global average temperatures well below 2\,$^\circ$C by 2100, with a goal of remaining below 1.5\,$^\circ$C. For Canada, this pledge translates into an 80\% reduction of net emissions from the 2005 level by 2050.$^{59}$ A 2018 IPCC (U.N.\  Intergovernmental Panel) report$^{7}$ demonstrates that 2$^\circ$C of warming, compared to 1.5$^\circ$C, poses significantly higher threats to physical, biological, and social systems, and risks irreversible impacts. 
Noting the Paris pledges are insufficient, the IPCC calls for 45\% global emissions cuts (from 2010 levels) by 2030, reaching {\em net zero} by 2050, to limit warming to 1.5$^\circ$C. Despite  widespread agreement on these goals, emissions are still rising today.$^{60}$ 

{\noindent\bf Social movements and calls to action:}  New
movements have emerged to call for action commensurate with the urgency of the situation.  
The student movement `FridaysForFuture', started by Greta Thunberg in August 2018,  is a prime example:  
on May 24, 2019, it inspired 1879 strikes across 133 countries, including thousands of Canadian students in 104 locations, while a Sept.\ 20, 2019 event involved over 3.2 million demonstrators at over 6000 events in over 165 countries.$^{61}$   Other groups, such as `Extinction Rebellion' and the `Sunrise Movement', have also mobilized rapidly and  significantly.  At the same time, a wave of climate-related litigation has been launched against governments and corporations,$^{62}$ bolstered recently by more precise scientific causal attribution of natural disasters.$^{63}$  Public  figures$^{64,65}$ as well as professional$^{66,67}$ and media$^{68}$ organizations, are urging swift action. 
All of Canada's major political parties recognize the reality of climate change.

\section{Astronomy's Role and Responsibility} \label{S:AstroRoleResponsibility}


At the most basic level, astronomy exists to serve humanity's quest for knowledge about its place and future in the Universe.
To ignore a clear and present threat to that place and that future would be irresponsible in the extreme.  Pursuits like astronomy would cease to exist if insufficient action is taken and the more alarming predictions materialize.  It is time for our professional community
to confront its climate impact and to plan for a low-carbon future.  Our long-range planning exercises provide ideal opportunities to declare our intentions. 

As one of the most public-facing of the sciences, astronomy must be especially clear and deliberate in its actions.  Astronomers have a responsibility to lead by example: our individual and collective actions must be consistent with our understanding of climate science and its implications.  Failure to act consistently degrades our  credibility,$^{69}$ and risks impairing public perception of the reality of the climate crisis and the need for action.$^{70}$ We must strive to acknowledge climate change whenever appropriate in our communications, outreach, and educational activities,$^{71}$ and to be reliable brokers of factual and scientific information when we do. 

Much of our climate impact arises from academic travel, contributing to the remarkable 5.7\%/yr {\em increase} in flight emissions over 2013-19. This is alarmingly out of step with the 6.4\%/yr {\em decrease} required to hit the IPCC's 2030 target.  Astronomers fly frequently for conferences, colloquia/seminars, committee meetings, collaborations, etc.  High mobility is generally valued and encouraged, and many believe it to be essential for professional success.
However, a study of UBC researchers$^{69}$ found no meaningful correlation between research productivity and travel.  (Salary {\em does} correlate, perhaps because the most senior researchers have the greatest travel emissions: up to 80 tons \cotwo/yr!)  Flying less is among the most effective ways that we can reduce our individual emissions.$^{72}$  When academic travel is essential, offsetting emissions with well-vetted carbon offset programs would be the environmentally responsible thing to do. 

Astronomical infrastructure such as telescopes, buildings, and computing facilities must also be counted toward our climate impact.  Since manufacture, transportation, construction, and operations all contribute to the impacts of our infrastructure, these must be considered, even for international projects.  (The SKA correlator, for example, is scoped$^{73}$ at an operational load of 2.3 MW, enough electricity$^{74}$ for $\sim$1080 homes.)

Communicating valid science to the public through education and outreach is core to astronomy's purpose; and while astronomers are not climate scientists, we do encounter closely related topics like planetary atmospheric evolution.  It is therefore imperative for us to remain well-enough informed to answer climate-related questions and misconceptions, and to refer to authoritative sources.  Our communications with governments, funding agencies, and academic institutions are also critically important, as our collective influence on policy decisions is not negligible.  Our input is crucial on policies that limit our own ability to act responsibly, such as those concerning whether the external costs of climate change can be considered in planning, or whether climate-related  costs like carbon offsets are legitimate research expenses.

\section{Calls to Action}

The astronomical community and its members can take individual, collective, and organized actions to respond to the climate crisis. Below, we outline several recommendations.  

\subsection{Reset expectations for travel and professional activity 
} \label{SS:ResetTravelExpectations}


  

To rise to the challenge, we must set new expectations that all academic travel (and resulting emissions) will be justified, reported, and reviewed, rather than encouraged or rewarded for its own sake. 
We need to shift to a culture in which limiting our carbon footprint 
is valued and incentivized. This can only be done with the support and engagement of all individuals, research institutions, universities, and funding agencies. We call on scientists, administrators, and managers of research at all levels to make this a priority and respond with the required large-scale structural transition within our profession.

At the level of individuals, this means keeping track of the carbon emissions of our professional activities and setting personal objectives to reduce them in line with {\cdm international targets and governmental commitments}. 
Air travel is the major contributor: a single round-trip transatlantic flight emits $\sim$1.6 tons of \cotwo\ per passenger.$^{72}$  This represents four months' worth of per-capita emissions at the current global average rate,$^1$ and over seven months' worth by 2030 if we adopt the IPCC guidelines for 1.5$^\circ$C.$^{7}$ In the analogous field of atmospheric research, one  study$^{75}$ found travel-related emissions of 4.7 tons \cotwo/yr per researcher.  At the institutional level, flying by faculty and staff is responsible for about a third of all carbon emissions from a typical university.$^{76}$

Air travel should therefore be carefully justified, factoring in the location and purpose of the event, and consideration should be given to options such as  remote participation, ground travel for distances $< 500$\,km, combining trips, and prioritising extended stays.  We need to resist our fear of missing out:
for example, one should choose to attend some large annual conferences every other year or when their geographical location is more convenient, and one could pass on trips that have lower benefits, such as invitations to give short presentations. {\cdm Because networking plays an important role during early career stages, graduate students and junior colleagues should be the preferred project representatives.}

{\cdm To be effective, these expectations must be adopted by institutions and granting agencies rather than simply reflecting individual attitudes. }
Many institutions and organisations have adopted a Code of Conduct supporting a low-carbon research culture.  The Tyndall Centre for Climate Change in the UK offers an excellent example$^{77}$ as it provides guidance for scientists to make decisions and track their emissions.

To amplify actions on a scale larger than the individual, institutional practices at the level of research institutions and universities should also shift to reduce academic air travel. Institutional policy should require justification for travel in proportion to the climate impact of the trip; for instance, a long-haul intercontinental flight should require far better justification than a shorter flight, which in turn should require more justification than ground transport. Institutions must track  travel-related emissions of their academics each year, comparing to pre-determined goals that aim for measurable reduction within rapid timescales, and revise their policy accordingly. Long-term monitoring is essential to ensure that initial efforts to reduce emissions are sustained, and to compare with others. The “Flying Less Policy” of Concordia University's Dept.\  of Geography, Planning \& Environment is an excellent example: it requires all faculty members to report annual flying activity, and publicly discloses$^{78}$ collective statistics. 

To remove barriers for individual actions, institutions must alter tenure and promotion expectations to remove incentives to fly, for example by re-wording appointments and guidelines to favour substantial written research contributions and high-impact presentations over a large number of comparatively low-impact presentations at international institutions.

When flying is approved as necessary, institutions must adopt an official policy wherein the associated carbon emissions are 100\% offset through reputable offset programs with measurable and verified outcomes (e.g.\ Less.ca, Carbonzero, Carbone Bor\'eal), and  must continually review their chosen offset programs.  Carbon offsets have been a controversial measure as they offer a limited response, are difficult to verify, and do not prevent carbon emissions in the first place; but they must still be pursued as a mitigating strategy for any essential flying.  

Granting agencies and governments, for their part, must recognize greenhouse gas emissions as true external costs of research activities; accordingly they should allow additional expenses for mitigation, such as the choice of more costly rail transport over the cheapest flight, or the purchase of carbon offsets.   Likewise, not choosing public transport option should require special justification in travel reimbursements.

When evaluating grant proposals,  considerations of estimates and mitigation of carbon emissions, sustainable strategies, and climate education initiatives should be required or rewarded, in the same way that the 
training of future researchers and equity, diversity and inclusivity have become merit factors. 


\subsection{Plan, monitor, and report impacts 
}\label{SS:PlanMonitorReport} 


In order to steadily reduce the carbon footprint of our profession year over year, Canadian astronomers and the organizations that employ them (university departments, observatories, planetariums, etc.) must become accustomed to framing their activities and operations in terms of carbon budgets. 
The logical first step would be a comprehensive survey of emissions from our professional activity,$^{79}$ organized by a CASCA Sustainability Committee (\S~\ref{SS:SustainabilityCommittee}) at its inauguration. 
Continuing from there, the budgeting process will ensure that we have reliable data to track our collective greenhouse gas emissions and are delivering on our commitment to become carbon-neutral.  For individuals, a number of carbon footprint calculators can be easily found online to inform them on what they should be tracking in their activities and what their choices add up to in equivalent CO$_2$ emissions.  Incentivizing this kind of process will be critical to its success; some ideas include: (1) organizations requiring their employees monitor and report their work-related CO$_2$ emissions as part of their own budgeting process; (2) granting agencies requiring PIs report their emissions as part of the ``Broader Impacts" section of their applications; and (3) rewarding PIs and groups who make significant progress towards reducing their GHG emissions and/or include climate science components in their courses.    





Organizations should strive for similar goals, by (1) tracking and reporting the carbon footprint of their activities and their staff and students; (2) ensuring that all simple means of reducing the carbon footprint are done by default (e.g., that computers suspend automatically; heating and cooling are minimized); (3) installing (or lobbying for) solar panels on buildings; (4) publishing greenhouse gas emissions and mitigation efforts on their web pages; (5) including experts on climate change or GHG reduction practices in their colloquia (e.g. once per semester); and (6) networking with other organizations on effective practices.  Large organizations like CASCA can hold or promote regular sessions, working groups, and collaboration space on climate change and its relationship to professional astronomy.

 

   

\subsection{Redesign astronomical conferences and colloquia  
}\label{SS:RedesignConferences}


%

Because astronomy research collaborations tend to be international, conferences are an important part of our professional lives. They are opportunities to learn of current research, and to interact with a diverse group of students, researchers and experts.  However, the environmental impact of in-person conferences requires us to explore and design alternatives,$^{80}$ such as online conferences or blended conferences.$^{81,82}$


\subsubsection{The potential of online conferences}\label{SSS:Confs-Online} 

Platforms developed for online teaching offer promise for conferences as well.  Acceptable systems show both the speaker (via webcam) and their presentation in real time, with options for uploading a prepared slide deck or for sharing the presenter's computer screen; provide tools to annotate slides during the presentation, or a whiteboard option (shared or presenter-only); allows allows participants to enter comments and questions via a chat feature, or ask questions via audio or video.  Integrated polling tools  allow the presenter to ask questions of the audience, and there is the possibility of having ``breakout rooms" for small group discussions and interactive white-boarding (i.e.\ collaboratively writing on a shared screen).  

Web conferencing platforms offer certain advantages over in-person presentations.  During online  presentations, for instance, participants can submit questions at any time via a chat feature, to be monitored by a session chair and highlighted at the end of the presentation.  This has the potential to involve a greater variety of people in the Q\&A period, as quieter and more junior members of our community may be more likely to participate.  A more traditional post-talk question period can be established using the hand-raising feature of the software and moderated by a session chair. All of this can be recorded for future reference, reducing unconscious biases in attribution. 

In addition to reducing the financial barrier to participation, online conferences can be more accessible to those with caregiver responsibilities, mobility limitations, or health problems that make travel inadvisable.  Online conferences would thus be more inclusive.  

Informal interactions and discussions are especially valued in the conference experience.   Online conferences can foster these using small group ``coffee break" discussion sessions around specific topics or themes, and encouraging one-on-one discussions. 
Posters can be available throughout the conference, with comments and questions submitted at any time, and  live poster sessions scheduled for discussions. 

Fully online conferences are possible. In February 2019, the Educational Developers Caucus (EDC) successfully ran an online conference with 255 participants, averaging 20-30 per session over four days.  
Attendance exceeded previous in-person EDC conferences, in great part due to the lower cost of participation. 

For our CASCA symposia, a blended model should be considered wherein attendees gather in groups in a few locations across the country to interact through teleconferencing, with others joining purely online. 


\subsubsection{Food and waste at conferences}\label{SSS:Confs-other}

{\cdm 
While the non-travel impacts of an individual conference are not large, they should not be ignored; and these relatively high-profile events can be indicative of the sincerity with which we approach climate mitigation. 

{\noindent \bf Food options:} Food production, especially land and energy-intensive meat and dairy production, has oversized climate impacts.  To quote a recent EAT-{\em Lancet} commission$^{83}$ on the future of food: 
{\small
\begin{addmargin}{0.5cm}
Food production is the largest cause of global environmental change. Agriculture occupies about 40\% of global land, and food production is responsible for up to 30\% of global greenhouse gas emissions and 70\% of freshwater use. Conversion of natural ecosystems to croplands and pastures is the largest factor causing species to be threatened with extinction. Overuse and misuse of nitrogen and phosphorus causes eutrophication and dead zones in lakes and coastal zones. Environmental burden from food production also includes marine systems. About 60\% of world fish stocks are fully fished, more than 30\% overfished, and catch by global marine fisheries has been declining since 1996.  
\end{addmargin}
}

This commission,
along with numerous other studies,$^{84-88}$ outlines major changes in food production and consumption patterns required if global climate, public health, and biodiversity are to become sustainable.
{
The impacts of these changes could be very significant. A report$^{11}$ of the IPCC states that ``shifting to diets that are lower in emissions-intensive foods like beef delivers a mitigation potential of $0.7-8.0$~Gt{\cotwo}eq/yr with most of the higher end estimates ($>$6 Gt{\cotwo}eq/yr) based on veganism, vegetarianism or very low ruminant meat consumption.'' Another study$^{85}$ found that a shift to diets  excluding animal products has the potential to reduce food’s land use by 76\%,  GHG emissions by 49\%, acidification by 50\%, and eutrophication by 49\%. The changes needed are significant: the diet recommended by the EAT-{\em Lancet} commission calls for a reduction in consumption of 84\% for red meat, 60\% for poultry and eggs, and 30\% for dairy products, compared with the current average North American diet.
}
With these points in mind, or for religious, health, or ethical reasons, many in the astronomical community eat a partially or completely plant-based diet. In our conferences, these choices are usually respected  with special meal options.  We recommend instead} 
that 
conference food be plant-based or at least vegetarian by default. This would appeal to the many non-vegetarians who make a point of reducing their meat consumption; currently, eating a meatless meal is not an option for those who did not register as vegetarians. In addition to decreasing the immediate consumption of meat, this would serve to normalize meatless eating and possibly inspire conference participants to 
shift dietary choices. 

{\noindent \bf Food waste:} While it is difficult to predict how much attendees will eat, we should work toward a shift in attitudes where what matters is that people get as much to eat as they need, not necessarily as much as they would indulge in if it is available. It should become acceptable and even laudable if all the food is eaten at a coffee break. The CASCA-2019 organizers set a good example by not having food at the afternoon break, only tea and coffee, which seemed sufficient in the afternoon.  While bylaws usually prevent donating excess food, care should be taken to ensure that it is composted. 

{\noindent\bf Non-food waste:} The amount of non-food waste is also an issue thanks to the use of plastics and other non-renewables, or the lack of proper recycling options. Various conference organizers have begun to tackle the problem.  {\cdm Simply choosing low-packaging providers or working with municipalities and conference centres to provide recycling and composting bins (with guidance on their use) would be significant steps. } Smaller interventions include encouraging attendees to bring their own mug and using compostable utensils.   A  plastic-free symposium of the Australian Marine Sciences Association (July 2019) demonstrates that further steps are possible. 

 
\subsubsection{Climate-friendly colloquia}\label{SSS:Colloquia}
Many of our recommendations for conferences and travel apply just as well to colloquia. 
Webinar software can be used for colloquia as it can for conferences, both for the talk itself and for less formal discussions.  In-demand speakers may be pleased to be able to reach more people while spending less time in airports and on planes.

Seminars and colloquia should be live streamed and archived as a matter of course.  This increases the reach of the presentation and normalizes the online seminar format: if scientists get used to watching seminars online and asking questions remotely, it becomes easier to progress towards a fully online seminar series.

When a speaker is brought in, ground travel should always be preferred and when flying is essential, carbon offsets should be purchased as a matter of course. In addition, efforts should be made to extend the speaker's stay for collaboration, and to coordinate with nearby institutions.

\subsection{Plan and create low-impact infrastructure 
 } \label{SS:Infrastructure} 


Beyond personal and systematic carbon output from professional activities, the infrastructure which we build, use, and replace contributes a significant fraction to our non-direct carbon footprint; in 2009, the global construction industry contributed 23\% of the total \cotwo\ emission produced by economic activity, 94\% of that emission being indirect.$^{89}$ In terms of physical infrastructure, the role astronomers can play in offsetting carbon comes not from our direct usage, but rather from our position of influence within our institutions: astronomers can advocate for carbon-neutral or carbon-negative infrastructure policies at an institutional level, making them a priority in the development of new and replacement infrastructure. Further, the profession's significant use of computing infrastructure makes consideration of its associated carbon footprint critical. In the case of computing, significant carbon and costs savings can be achieved with relatively simple shifts in computing habits. 

\noindent \textbf{Low-Carbon Construction:} Given the significant carbon footprint of physical infrastructure, 
low-carbon methods of construction become critical. This is particularly important 
in the developing world, where 
extensive infrastructure plans would, if built with standard techniques, expend 
35-60\% of the global carbon budget for 2050.$^{90}$ Given the long time span of infrastructure projects,  fully implementing current sustainability best practices, along with the development of new technology, becomes critical to avoiding these emissions.  Architecture associations' climate declarations and commitments$^{66}$ provide guidance on how this should be accomplished. 

One promising new technology 
is {\em mass timber}, which uses engineered wood products, such as cross-laminated  or glue-laminated timber, to replace or augment steel- or concrete-based structural components,  avoiding emissions from the manufacture of steel and concrete, while reducing weight and 
sequestering carbon in the structural material.$^{91,92}$ 
It is estimated$^{93}$ that the use of wood-based materials could save 14-31\% of total global carbon emissions, when indirect emissions and carbon sequestration are included.  Canada has become a leader in their development and adoption.$^{94,95}$
Academic institutions, with sufficient risk tolerance and 
means to impact policy at the municipal level, are already planning or have completed carbon-negative mass timber buildings, including at the University of British Columbia$^{96}$ and the University of Toronto.$^{97}$ Astronomers in campus settings can lobby universities to consider these technologies while adopting  carbon-negative building policies. 

\noindent\textbf{
Cloud Computing:} 
Astronomers have  ever-increasing computing requirements, traditionally met by 
individual researchers' machines or clusters. This computing infrastructure is often under-utilized and inefficient, leading to significant direct and indirect carbon emission when operations, production and transport are considered.   Cloud computing has the potential to provide sufficient computing power with significantly reduced carbon footprint; emissions can be reduced by a factor of nine by shifting to a public cloud.$^{98}$ Cloud computing providers can optimize against carbon emission, with most public providers already carbon neutral and committed to fully renewable power in the future.$^{99}$
Cloud computing can also be more cost-effective. 
Importantly, a transition to cloud computing can be accomplished very rapidly and with almost no change to the users' infrastructure. 


\subsection{Advocate divestment from fossil fuels and their financing}\label{SS:divest} 
Through their investment choices, our institutions' endowments and pension funds exercise far greater influence on global carbon budget than we do individually, and while 8136 universities and colleges are signatories to climate emergency declarations$^{100,101}$ many remain invested in fossil fuel extraction and the banks and  asset management companies$^{102}$ that finance them.  Continued pressure and increasing recognition of financial risks have however led to significant successes,  such as the recent divestment of US\$80 billion by the University of California system.$^{103}$ 


\section{Climate Science in Education and Outreach 
\label{S:EducationOutreach} 
} 


Astronomers understand the basics of atmospheres, radiative transfer, and convection, making us relatively well-qualified to talk about climate physics.  Moreover, astronomers are among the scientists most frequently in direct contact with the general public, and we reach many non-scientists with introductory astronomy (Astro-101). 

At public events and in our classes, we are often expected to field very wide-ranging questions, including those about climate.  In Canada, the literacy rate is $\sim$99\%, but the science literacy rate$^{104}$ is 42\%. 
Indeed, we are often faced with  misconceptions and sophistries concerning climate. Given this, it is essential that astronomers understand common misconceptions and misrepresentations related to climate change, and be aware of the key scientific findings.  

Astro-101 is the only university-level science course many students will ever take, and their introduction to how science works.  Although climate is not the course's direct object,  many Astro-101 topics, such as the properties of  Solar System bodies, connect naturally to climate science.
These are crucial opportunities to equip students with some of the knowledge and skills needed to understand and evaluate climate science and climate change.

All introductory astronomy courses include discussions of the electromagnetic spectrum, spectroscopy, and atmospheric absorption: basic elements of the greenhouse mechanism as well as the foundations of our field. 
Planetary habitability -- be it of Solar System worlds or exoplanets -- is intimately tied to climate change: the outer (cold) edge of the habitable zone relates to the faint young Sun paradox, while the inner edge relates to the runaway greenhouse.   Earth's climate sensitivity therefore has broad connections to the topic of habitability.  {\cdm Indeed, the human dimensions of climate change make such scientific topics directly relevant to students' lives, so highlighting these connections is not just our responsibility, but good pedagogy.}

In our opinion, a crucial lesson to impart is that the climate crisis, or any human crisis for that matter, cannot ever be avoided by space travel to some off-world colony, even if these exist -- and that our searches for Earth twins and habitable environments are not for this purpose.  There is, in short, no `Planet B'.$^{105}$  Students must learn, instead, of  Earth's unique status as an inhabited planet and come to appreciate humanity's duty to keep it that way. 

A curated collection of resources related to climate science and to the nature of science would be very beneficial. This would include both factual information based on peer-reviewed studies (review articles, statistics, infographics) and pedagogical tools such as case studies, lecture-tutorials, and interactive animations. Creating such a resource would involve verifying external sources and creating tailored material particularly appropriate for Canadian astronomers and astronomy educators. 
We therefore call for the establishment of such a resource, perhaps by a CASCA sustainability committee (\S\,\ref{S:RoleOfCASCA}).  
In addition, we endorse the {\em Astro2020} white paper $^{71}$ on these issues
by Williamson et al.

\section{The Role of CASCA} \label{S:RoleOfCASCA} 

\subsection{Establish sustainability committee 
} \label{SS:SustainabilityCommittee} 




We must (1) clearly acknowledge that the link between human activity and climate change is firmly established scientifically, that there is an urgent need for a strong response to the climate crisis, and that current efforts by governing agencies are insufficient, (2) declare our common goal to act rapidly in transforming our professional practices to minimize their impact on the environment, in particular by reducing our net carbon emissions in accordance with the IPCC targets, and (3) establish a sustainability committee to develop and start implementing an effective course of actions to meet our goals. Other scientific communities worldwide have already started on this path.$^{106,107}$  For instance, the AAS has endorsed the American Geophysical Union's statement on ``compelling evidence of human impact on the climate system with potentially far-reaching consequences for ecological and political systems".  It has also established a sustainability committee that is charged with the review and recommendation of plans related to energy conservation within the AAS and ``fostering awareness and participatory social responsibility for all AAS members".  It is time we join them.

Designing a holistic climate action plan for Canadian astronomy will be the responsibility of our sustainability committee.  This committee would also be responsible for monitoring and reporting the impacts of our actions. Our actions and their results should be communicated broadly and regularly through the committee's website, and when appropriate, cited in communications to governments, funding agencies, media and the public.  In many areas, official policies will be needed to implement and facilitate more sustained actions, and this can only be done with support from government, research institutions, universities, and funding agencies.  It will be our sustainability committee's responsibility to engage in and sustain the negotiations that will ensure this support.

\subsection{Establish a de-carbonization roadmap for Astronomy 
}



 

To effect significant and long-lasting change, we must  develop and adhere to a detailed roadmap for reduction in greenhouse gases, in accordance with the IPCC global recommendations. Through its Sustainability Committee, CASCA should establish such a roadmap and organize periodic reviews of our progress.  The roadmap will be most valuable if endorsed by CASCA, member institutions, and possibly other major institutions such as NSERC. It should also be subject to amendment in light of new research.

Some examples of institutional climate action plans include those of the University of Calgary and Queen's University.$^{108}$ These  detail the actions done by the institutions to date and lay out clear emissions reductions goals for the future, divided into different areas such as transportation and infrastructure. Along with each emissions goal, these documents lay out an implementation strategy including costs and oversight. A similar style of action plan produced and tailored for the field of Astronomy would help make the goals laid out in this paper achievable. 

Importantly, targets set for Astronomy should be technically viable and  never invoke `magic wands': potential future technologies which have not yet been developed or commercialized. If possible, we recommend the establishment of an action plan to be done with the help of external experts and professional planners.  The document detailing this road-map should be thorough, visually clear and compelling, easily accessible, and distributed to Canadian astronomers.

\subsection{Advocate with Granting Agencies and Institutions
} \label{SS:AdvocateWithAgencies} 


A significant role for CASCA, as for the Long Range Plan, is to convey to governments and granting agencies the community consensus on climate action, and to highlight specific changes to enable the actions we wish to take. 
For one, climate and environmental impacts must be considered real external costs of research, and for this reason mitigation expenses like carbon offset costs and more expensive rail transport, should be allowed reimbursements.   For another, granting agencies should establish and reinforce expectations around the reporting of research-related climate impacts, as we highlighted in \S\ref{SS:ResetTravelExpectations}. 

Most important, however, is that governmental leadership will be required to ration greenhouse gas emissions, ensuring that our discipline, and all of Canada, do not exceed their shares of the global carbon budget.  CASCA has a critical role to play in conveying this message to government bodies. 




\section{Concluding remarks }\label{S:Conclusions} 

Astronomers are uniquely prepared to appreciate the vast magnitude of the global climate crisis and its implications for the future of our only known inhabited planet. 
Astronomy must, therefore, like every field of human activity, take urgent steps to mitigate the developing climate crisis and avoid the worst potential outcomes while adapting to those that are inevitable.  
For this reason, we recommend that climate responsibility be an explicit priority in the 2020 {\em Long Range Plan}, and that it be used to guide all aspects of our profession, from our research and education practices to our infrastructure. 




\noindent {\bf \Large{References}} \\ 
\begin{itemize}
\item[1.] \href{https://ourworldindata.org/co2-and-other-greenhouse-gas-emissions}{Our world in data: CO2 and greenhouse gas emissions} \url{https://ourworldindata.org/co2-and-other-greenhouse-gas-emissions}. Accessed: 2019-09-06.
\item[2.] Siegenthaler, U. et al. Stable carbon cycle-climate relationship during the late Pleistocene. {\em Science} 310, 1313-1317 (2005).
\item[3.] NOAA Earth System Research Laboratory \url{https://www.esrl.noaa.gov/}. Accessed: 2019-07-31.
\item[4.] Schurer, A. P., Mann, M. E., Hawkins, E., Tett, S. F. \& Hegerl, G. C. Importance of the pre-industrial baseline for likelihood of exceeding Paris goals. {\em Nature climate change} 7, 563 (2017).
\item[5.] Meehl, G. A. et al. How much more global warming and sea level rise? {\em Science} 307, 1769-1772 (2005).
\item[6.] Pachauri, R. K. et al. Climate change 2014: synthesis report. Contribution of Working Groups I, II and III to the fifth assessment report. Intergovernmental Panel on Climate Change 151 (2014).
\item[7.] Intergovernmental Panel on Climate Change. Global Warming of 1.5$^\circ$ C: An IPCC Special Report on the Impacts of Global Warming of 1.5$^\circ$ C Above Pre-industrial Levels and Related Global Greenhouse Gas Emission Pathways, in the Context of Strengthening the Global Response to the Threat of Climate Change, Sustainable Development, and Efforts to Eradicate Poverty (IPCC, 2018).
\item[8.] Fischer, H. et al. Palaeoclimate constraints on the impact of 2 C anthropogenic warming and beyond. {\em Nature geoscience} 11, 474 (2018).
\item[9.] World Climate Research Program: CMIP6 Model Analysis Workshop Summary \url{https://www.wcrp-climate.org/news/wcrp-news/1478-cmip6-first-results}. Accessed: 2019-07-31.
\item[10.] Natural Resources Canada. Canada's Changing Climate Report \url{https://changingclimate.ca/CCCR2019}. Accessed: 2019-07-31.
\item[11.] Intergovernmental Panel on Climate Change. Climate Change and Land: An IPCC special report on climate change, desertification, land degradation, sustainable land management, food security, and greenhouse gas fluxes in terrestrial ecosystems (IPCC, 2019).
\item[12.] United Nations IPCC. Special Report on the Ocean and Cryosphere in a Changing Climate \url{https://www.wsj.com/articles/the-university-of-california-divests-11569188481}. Accessed: 2019-09-25. 2019.
\item[13.] Wuebbles, D. J., Fahey, D. W. \& Hibbard, K. A. Climate science special report: fourth national climate assessment, volume I (2017).
\item[14.] Coats, D. R. Statement for the Record: Worldwide Threat Assessment of the US Intelligence Community, Daniel R. Coats, Director of National Intelligence, Senate Select Committee on Intelligence, January 29, 2019 in United States. Office of the Director of National Intelligence; United States. Congress. Senate. Select Committee on Intelligence (2019).
\item[15.] State of the Climate 2018 \url{https://www.csiro.au/en/Showcase/state-of-the-climate}. Accessed: 2019-08-19.
\item[16.] Climate Emergency Declaration and Mobilisation in Action \url{https://www.cedamia.org/global/}. Accessed: 2019-09-25.
\item[17.] Box, J. E. et al. Global sea-level contribution from Arctic land ice: 1971-2017. {\em Environmental Research Letters} 13, 125012 (2018).
\item[18.] Parkinson, C. L. A 40-y record reveals gradual Antarctic sea ice increases followed by decreases at rates far exceeding the rates seen in the Arctic. {\em Proceedings of the National Academy of Sciences} 116, 14414-14423 (2019).
\item[19.] Farquharson, L. M. et al. Climate change drives widespread and rapid thermokarst development in very cold permafrost in the Canadian High Arctic. {\em Geophysical Research Letters} (2019).
\item[20.] Maurer, J., Schaefer, J., Rupper, S. \& Corley, A. Acceleration of ice loss across the Himalayas over the past 40 years. {\em Science advances} 5, eaav7266 (2019).
\item[21.] Turetsky, M. R. et al. Permafrost collapse is accelerating carbon release. {\em Nature} 569, 32 (2019).
\item[22.] Sweet, W. et al. NOAA Technical Report NOS CO-OPS 083: Global and regional sea level rise scenarios for the United States. Silver Spring, MD (2017).
\item[23.] Dumitru, O. A. et al. Constraints on global mean sea level during Pliocene warmth. {\em Nature} (2019).
\item[24.] Nadi Bay Declaration on the Climate Change Crisis in the Pacific \url{https://cop23.com.fj/nadi- bay-declaration-on-the-climate-change-crisis-in-the-pacific/}. Accessed: 2019-08-16.
\item[25.] Sharma, E. et al. in The Hindu Kush Himalaya Assessment 1-16 (Springer, 2019).
\item[26.] Yeung, A. C., Paltsev, A., Daigle, A., Duinker, P. N. \& Creed, I. F. Atmospheric change as a driver of change in the Canadian boreal zone1. {\em Environmental Reviews}, 1-31 (2018).
\item[27.] Streletskiy, D. A., Suter, L. J., Shiklomanov, N. I., Porfiriev, B. N. \& Eliseev, D. O. Assessment of climate change impacts on buildings, structures and infrastructure in the Russian regions on permafrost. {\em Environmental Research Letters} 14, 025003 (2019).
\item[28.] Schiermeier, Q. Climate change made Europe's mega-heatwave five times more likely. {\em Nature} 571, 155 (2019).
\item[29.] Marvel, K. et al. Twentieth-century hydroclimate changes consistent with human influence. {\em Nature} 569, 59 (2019).
\item[30.] 17 Countries, Home to One-Quarter of the World's Population, Face Extremely High Water Stress \url{https://www.wri.org/blog/2019/08/17-countries-home-one-quarter-world-population-face-extremely-high-water-stress}. World Resources Institute. Accessed: 2019-08-19.
\item[31.] Abatzoglou, J. T. \& Williams, A. P. Impact of anthropogenic climate change on wildfire across western US forests. {\em Proceedings of the National Academy of Sciences} 113, 11770-11775 (2016).
\item[32.] Aumann, H. H., Behrangi, A. \& Wang, Y. Increased Frequency of Extreme Tropical Deep Convection: AIRS Observations and Climate Model Predictions. {\em Geophysical Research Letters} 45, 13-530 (2018).
\item[33.] Mora, C. et al. Broad threat to humanity from cumulative climate hazards intensified by greenhouse gas emissions. {\em Nature Climate Change} 8, 1062 (2018).
\item[34.] Mills, E. Insurance in a climate of change. {\em Science} 309, 1040-1044 (2005).
\item[35.] Hoegh-Guldberg, O. et al. The human imperative of stabilizing global climate change at 1.5 C. {\em Science} 365, eaaw6974 (2019).
\item[36.] Diaz, S. et al. Summary for policymakers of the global assessment report on biodiversity and ecosystem services of the Intergovernmental Science-Policy Platform on Biodiversity and Ecosystem Services (2019).
\item[37.] Great Barrier Reef Marine Park Authority. Great barrier reef outlook report 2019 (2019).
\item[38.] Board, O. S., Council, N. R., et al. Ocean acidification: a national strategy to meet the challenges of a changing ocean (National Academies Press, 2010).
\item[39.] Barrett, M. et al. Living planet report 2018: Aiming higher (2018).
\item[40.] Rosenberg, K. V. et al. Decline of the North American avifauna. {\em Science} (2019).
\item[41.] Pimm, S. L. et al. The biodiversity of species and their rates of extinction, distribution, and protection. {\em Science} 344, 1246752 (2014).
\item[42.] World Health Organization. COP24 special report: health and climate change (2018).
\item[43.] Crowley, R. A. Climate change and health: a position paper of the American College of Physicians. Annals of internal medicine 164, 608-610 (2016).
\item[44.] U.S. Call to Action on Climate, Health, and Equity: A Policy Action Agenda \url{https://climatehealthaction.org/cta/climate-health-equity-policy/}. Accessed: 2019-09-06.
\item[45.] AMA Position Statement on Climate Change and Human Health \url{https://ama.com.au/media/climate-change-health-emergency}.Accessed: 2019-09-06.
\item[46.] McMichael, A. J., Woodruff, R. E. \& Hales, S. Climate change and human health: present and future risks. {\em The Lancet} 367, 859-869 (2006).
\item[47.] Watts, N. et al. The Lancet Countdown on health and climate change: from 25 years of inaction to a global transformation for public health. {\em The Lancet} 391, 581-630 (2018).
\item[48.] Wheeler, T. \& Von Braun, J. Climate change impacts on global food security. {\em Science} 341, 508-513 (2013).
\item[49.] Bathiany, S., Dakos, V., Scheffer, M. \& Lenton, T. M. Climate models predict increasing temperature variability in poor countries. Science Advances 4. eprint: \url{https://advances.sciencemag.org/ content/4/5/eaar5809.full.pdf}. \url{https://advances.sciencemag.org/content/ 4/5/eaar5809} (2018).
\item[50.] Watkins, K. et al. Human Developmnent Report 2007/2008 - Fighting climate change: Human solidarity in a divided world. \url{http://hdr.undp.org/sites/default/files/hdr_20072008_summary_english.pdf} (2008).
\item[51.] Programme, U. N. D. Human Development Report 1998 146. \url{https://www.un-ilibrary.org/content/publication/bc57a94d-en} (1998).
\item[52.] World Bank data: CO2 emissions (metric tons per capita) \url{https://data.worldbank.org/indicator/en.atm.co2e.pc}. Accessed: 2019-09-04.
\item[53.] Byers, E. et al. Global exposure and vulnerability to multi-sector development and climate change hotspots. {\em Environmental Research Letters} 13, 055012 (2018).
\item[54.] Rigaud, K. et al. Groundswell: Preparing for Internal Climate Migration (Washington, DC: World Bank) 2018.
\item[55.] Mach, K. J. et al. Climate as a risk factor for armed conflict. {\em Nature}, 1 (2019).
\item[56.] Watt-Cloutier, S. Petition to the Inter American Commission on Human Rights seeking relief from violations resulting from global warming caused by acts and omissions of the United States. Inuit Circumpolar Council Canada (2005).
\item[57.] Green, D., King, U. \& Morrison, J. Disproportionate burdens: the multidimensional impacts of climate change on the health of Indigenous Australians. {\em Medical Journal of Australia} 190, 4 (2009).
\item[58.] On Climate Change Risks, T. E. P. \& Potential, A. Canada's Top Climate Change Risks. Council of Canadian Academies (2019).
\item[59.] Canada's Mid-Century Long-Term Low-Greenhouse Gas Development Strategy \url{http://unfccc.int/files/focus/long-term_strategies/application/pdf/canadas_mid-century_long-term_strategy.pdf}.
\item[60.] Le Qu\'er\'e, C. et al. Global Carbon Budget 2018. Earth System Science Data 10, 2141-2194. \url{https: //www.earth-syst-sci-data.net/10/2141/2018/} (2018).
\item[61.] Newsletter For Future, Sept 21, 2019 \url{https://www.fridaysforfuture.org/news}. Accessed: 2019-09-21.
\item[62.] Sabin Center for Climate Change Law. Climate Change Litigation Databases \url{http://climatecasechart.com/}. Accessed: 2019-09-22. 2019.
\item[63.] Burger, M., Horton, R. \& Wentz, J. The Law and Science of Climate Change Attribution. {\em Columbia Journal of Environmental Law}, forthcoming (2020).
\item[64.] Pope Francis. Encyclical on climate change and inequality: On care for our common home (Melville House, 2015).
\item[65.] Prince Charles: We must act on climate change to avoid `potentially devastating consequences' \url{https://www.telegraph.co.uk/news/2017/01/22/prince-charles-must-act-climate-change-avoid-potentially-devastating/}. Accessed: 2019-09-17.
\item[66.] Architects' Declaration on Climate Change and Biodiversity: UK and Australia \url{https://www.architectsdeclar com}, \url{https://www.architectsdeclare.com.au}. Accessed: 2019-09-06.
\item[67.] Scientists' Warning on Climate Change \url{https://scientistswarning.forestry.oregonstate. edu/}. Accessed: 2019-09-17.
\item[68.] Covering Climate Now Project \url{https://www.coveringclimatenow.org}. Accessed: 2019-09-17.
\item[69.] Wynes, S., Donner, S. D., Tannason, S. \& Nabors, N. Academic air travel has a limited influence on professional success. {\em Journal of Cleaner Production} 226, 959-967 (2019).
\item[70.] Attari, S. Z., Krantz, D. H. \& Weber, E. U. Statements about climate researchers' carbon footprints affect their credibility and the impact of their advice. {\em Climatic Change} 138, 325-338. ISSN: 1573-1480. \url{https: //doi.org/10.1007/s10584-016-1713-2} (Sept. 2016).
\item[71.] Williamson, K., Rector, T. A. \& Lowenthal, J. Embedding Climate Change Engagement in Astronomy Education and Research. arXiv:1907.08043 (2019).
\item[72.] Wynes, S. \& Nicholas, K. A. The climate mitigation gap: education and government recommendations miss the most effective individual actions. {\em Environmental Research Letters} 12, 074024 (2017).
\item[73.] Ford, D., Faulkner, A., Kim, J. \& Alexander, P. SKA Software Correlator Concept Description (2011).
\item[74.] EPA Greenhouse Gas Equivalencies Calculator \url{https://www.epa.gov/energy/greenhouse-gas-equivalencies-calculator}. Accessed: 2019-09-06.
\item[75.] Stohl, A. The travel-related carbon dioxide emissions of atmospheric researchers. {\em Atmospheric Chemistry and Physics} 8, 6499-6504 (2008).
\item[76.] UC Santa Barbara climate action plan 2016 \url{http://www.sustainability.ucsb.edu/wp-content/uploads/Draft_2016-CAP_2_1_2017.pdf}.
\item[77.] Tyndall Centre for Climate Change Research Travel Strategy \url{https://tyndall.ac.uk/travel-strategy}. Accessed: 2019-09-21.
\item[78.] Concordia University's Declaration of Climate Emergency \url{https://www.concordia.ca/artsci/geography-planning-envionment/climate-emergency.html}. Accessed: 2019-09-22.
\item[79.] e.g., The Keystone Project: Carbon Footprint: Air Travel at the University of Washington \url{https://green.uw.edu/sites/default/files/keystoneproject_update.pdf}. Accessed: 2019- 09-22. 2014.
\item[80.] Desiere, S. The carbon footprint of academic conferences: Evidence from the 14th EAAE Congress in Slovenia. {\em EuroChoices} 15, 56-61 (2016).
\item[81.] Hamant, O., Suanders, T. \& Viasnoff, V. Seven steps to make travel to scientific conferences more sustainable. {\em Nature} 573, 451-452 (2019).
\item[82.] Middleton, C. Environmental sustainability goals drive changes in conference practices. {\em Physics Today} 72, 29 (9 2019).
\item[83.] Willett, W. et al. Food in the Anthropocene: the EAT-Lancet Commission on healthy diets from sustainable food systems. {\em The Lancet} 393, 447-492 (2019).
\item[84.] Springmann, M. et al. Options for keeping the food system within environmental limits. {\em Nature} 562, 519 (2018).
\item[85.] Poore, J. \& Nemecek, T. Reducing food's environmental impacts through producers and consumers. {\em Science} 360, 987-992. ISSN: 0036-8075. eprint: \url{https://science.sciencemag.org/content/360/ 6392/987.full.pdf}. \url{https://science.sciencemag.org/content/360/6392/987} (2018).
\item[86.] Greenpeace International. Less is more: reducing meat and dairy for a healthier life and planet (Greenpeace, 2018).
\item[87.] Food and Land Use Coalition. Growing Better: Ten Critical Transitions to Transform Food and Land Use. The Global Consultation Report of the Food and Land Use Coalition. (Food and Land Use Coalition, \url{https://www.foodandlandusecoalition.org}, 2019).
\item[88.] Kim, B. F. et al. Country-specific dietary shifts to mitigate climate and water crises (2019).
\item[89.] Huang, L., Krigsvoll, G., Johansen, F., Liu, Y. \& Zhang, X. Carbon emission of global construction sector. {\em Renewable and Sustainable Energy Reviews} 81, 1906-1916. ISSN: 1364-0321. \url{http://www.sciencedirect. com/science/article/pii/S1364032117309413} (2018).
\item[90.] M\"uller, D. B. et al. Carbon Emissions of Infrastructure Development. {\em Environmental Science \& Technology} 47. PMID: 24053762, 11739-11746. eprint: \url{https://doi.org/10.1021/es402618m}. \url{https: //doi.org/10.1021/es402618m} (2013).
\item[91.] Cornwall, W. Would you live in a wooden skyscraper? {\em Science}. \url{https://doi.org/10.1126/science.aah7334} (Sept. 2016).
\item[92.] Tollefson, J. The wooden skyscrapers that could help to cool the planet. {\em Nature} 545, 280-282. \url{https: //doi.org/10.1038/545280a} (May 2017).
\item[93.] Oliver, C. D., Nassar, N. T., Lippke, B. R. \& McCarter, J. B. Carbon, Fossil Fuel, and Biodiversity Mitigation With Wood and Forests. Journal of Sustainable Forestry 33, 248-275. eprint: \url{https://doi.org/10. 1080/10549811.2013.839386}.  (2014).
\item[94.] Urban One Builders \url{http://urbanonebuilders.com/}. Accessed: 2019-09-23.
\item[95.] Mass Timber Institute \url{https://www.masstimberinstitute.ca/}. Accessed: 2019-09-23.
\item[96.] UBC opens North America's greenest building \url{https://sustain.ubc.ca/news/ubc-opens-north-americas-greenest-building}. Accessed: 2019-09-23.
\item[97.] U of T to build academic wood tower on downtown Toronto campus \url{https://www.utoronto.ca/news/u-t-build-academic-wood-tower-downtown-toronto-campus}. Accessed: 2019- 09-23.
\item[98.] Is Cloud Computing Always Greener \url{https://www.nrdc.org/sites/default/files/cloud-computing-efficiency-IB.pdf}. Accessed: 2019-09-23.
\item[99.] How Green is your cloud \url{https://www.infoworld.com/article/3015632/how-green-is-your-cloud.html}. Accessed: 2019-09-23.
\item[100.] Higher and Further Education Institutions across the globe declare Climate Emergency \url{https://www.unenvironment.org/news-and-stories/press-release/higher-and-further-education-institutions-across-globe-declare}.Accessed: 2019-08-03.
\item[101.] Universities and Colleges for the Climate Summit \url{https://www.sdgaccord.org/climateletter}. Accessed: 2019-08-03.
\item[102.] Rainforest Action Network. Banking on Climate Change \url{https://www.ran.org/bankingonclimatechange2} Accessed: 2019-09-25. 2019.
\item[103.] Wall Street Journal Editorial Board. The University of California Divests. \url{https://www.wsj.com/ articles/the-university-of-california-divests-11569188481}. Accessed: 2019-09- 25. 2019.
\item[104.] Council of Canadians. Science Culture: Where Canada Stands \url{https://cca-reports.ca/reports/ science-culture-where-canada-stands}. Accessed: 2019-09-24. 2014.
\item[105.] Rees, M. On the future: Prospects for humanity (Princeton University Press, 2018).
\item[106.] No Fly Climate Sci \url{https://noflyclimatesci.org}. Accessed: 2019-09-24.
\item[107.] Labos 1.5 \url{https://labos1point5.org/en/home/}. Accessed: 2019-09-24.
\item[108.] University of Calgary and Queen's University's Climate Action Plans \url{https://www.ucalgary.ca/ sustainability/files/sustainability/climate-action-plan_final.pdf}, \url{https://www.queensu.ca/sustainability/campus-initiatives/climate-action/climate-action-plan}. Accessed: 2019-09-24.
\end{itemize} 


\end{document}